\documentclass[final]{anthology-ch}         

\usepackage[backend=biber, style=numeric]{biblatex}

\usepackage{booktabs}
\usepackage{graphicx}
\usepackage{float}
\usepackage{hyperref}
\usepackage{silence}
\title{Moving Pictures of Thought: Extracting Visual Knowledge in Charles S. Peirce’s Manuscripts with Vision-Language Models}

\author[1]{Carlo Teo Pedretti}[
  orcid=0000-0002-4115-0078
]

\author[2]{Davide Picca}[
  orcid=0000-0003-2014-0855
]

\author[3]{Dario Rodighiero}[
  orcid=0000-0002-1405-7062
]

\affiliation{1}{Department of Classics, University Sapienza of Rome, Rome, Italy}
\affiliation{2}{Department of Language and Communication Sciences, University of Lausanne, Lausanne, Switzerland}
\affiliation{3}{Campus Fryslân, University of Groningen, Groningen, The Netherlands}


\pubyear{2025}
\pubvolume{1}
\pagestart{1}
\pageend{1}
\conferencename{Proceedings of Conference XXX}
\conferenceeditors{Editor1 Editor2}
\doi{00000/00000} 

\begin{document}

\maketitle

\begin{abstract}
Diagrams are crucial yet underexplored tools in many disciplines, demonstrating the close connection between visual representation and scholarly reasoning. However, their iconic form poses obstacles to visual studies, intermedial analysis, and text-based digital workflows. In particular, Charles S. Peirce consistently advocated the use of diagrams as essential for reasoning and explanation. His manuscripts, often combining textual content with complex visual artifacts, provide a challenging case for studying documents involving heterogeneous materials. In this preliminary study, we investigate whether Visual Language Models (VLMs) can effectively help us identify and interpret such hybrid pages in context. First, we propose a workflow that (i) segments manuscript page layouts, (ii) reconnects each segment to IIIF-compliant annotations, and (iii) submits fragments containing diagrams to a VLM. In addition, by adopting Peirce’s semiotic framework, we designed prompts to extract key knowledge about diagrams and produce concise captions. Finally, we integrated these captions into knowledge graphs, enabling structured representations of diagrammatic content within composite sources.
\end{abstract}

\section{Introduction}

Diagrams play a central role in many forms of reasoning, from mathematics to philosophy and religious art~\cite{latour1990visualisation, rodighiero2023iiif, hamburger2019devotion}. Among the most prominent theorists of diagrammatic reasoning is Charles S. Peirce, who conceived of diagrams as a subtype of icon capable of representing and manipulating internal structures through visual means~\cite{dewaal2013perplexed, stjernfelt2007diagrammatology}. In his unpublished manuscripts, diagrams such as \textit{Existential Graphs} illustrate logical inferences via spatial configurations, offering a visual alternative to symbolic logic. Peirce referred to these constructs as “moving pictures of thought”~\cite{peirceCP16} (\textit{Collected Papers} 4.8), underlining their dynamic and epistemological function.

This idea finds material expression in Peirce’s manuscripts, where textual content, visual artifacts, and complex layouts are seamlessly integrated~\cite{keeler2020a_hidden}. These documents reflect both his theoretical commitment to diagrammatic reasoning and its practical development through layered and visually structured writing. However, this visual richness remains largely inaccessible in existing printed editions, which are compiled under severe editorial constraints~\cite{keeler1998iconic, keeler2020b_archive}.

Building on recent advances in the textual analysis of Peirce’s manuscript \textit{Prolegomena to an Apology for Pragmaticism} (PAP)~\cite{picca2023peirce}, we extend the investigation to his visual thinking. This exploratory study investigates the extent to which Vision Language Models (VLMs) can engage with diagrams as operative semiotic forms. To address this question, we propose a flexible and interoperable workflow for extracting structured knowledge from multimodal documents that supports integration within Linked Open Data (LOD) environments. Our modular workflow begins with the segmentation of page layouts to isolate diagrams, which are then linked IIIF annotations. These fragments are submitted to a VLM via prompt-based interactions informed by Peirce’s semiotic theory, with the aim of generating structured descriptions of diagrammatic content. The resulting outputs were then serialized in RDF, enabling machine-readable representations of visual reasoning within complex multimodal sources.

\begin{figure}[htbp]
  \centering
  \includegraphics[height=0.55\linewidth]{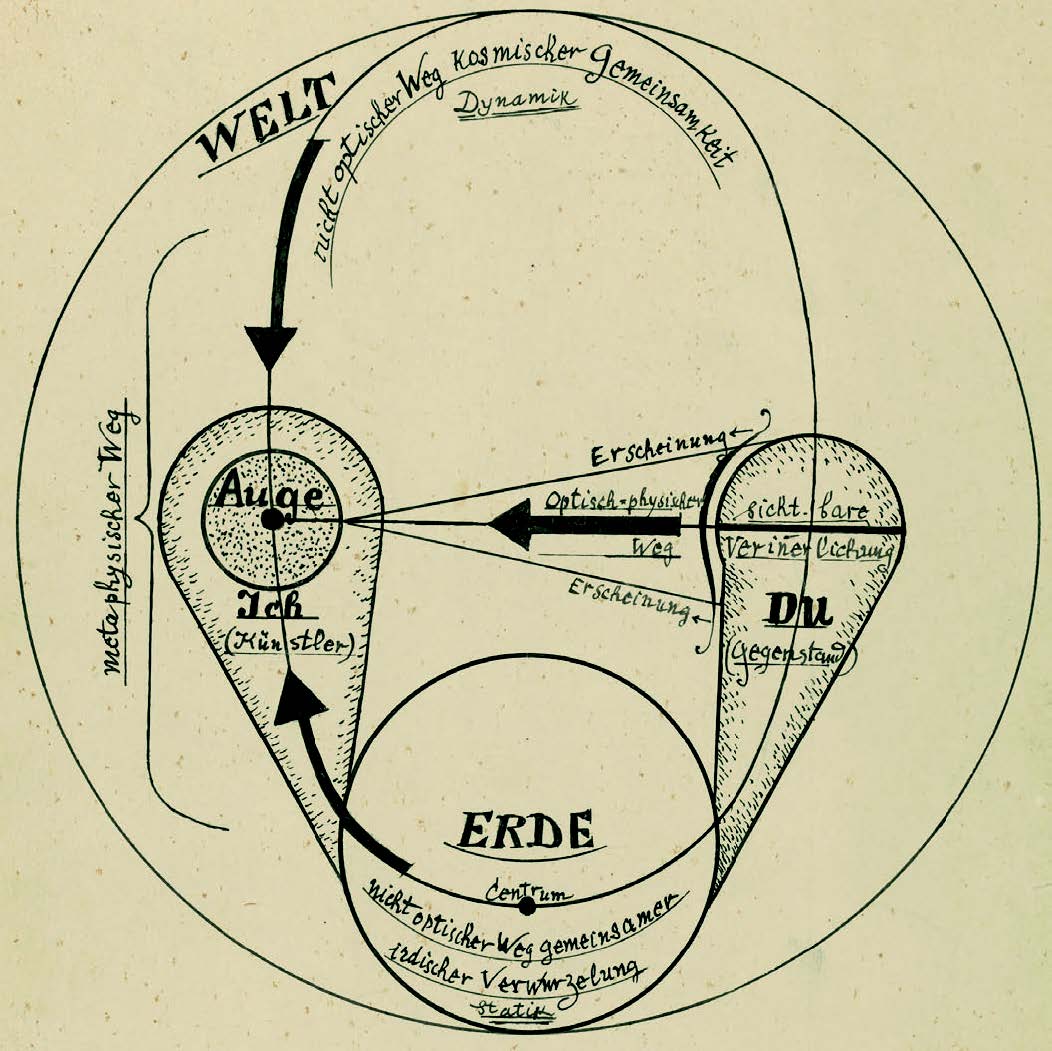}
  \caption{Paul Klee's Theory of Pictorial Configuration as a diagram. Zentrum Paul Klee, Bern, Inv.Nr. BG A/030. Photo: Zentrum Paul Klee.}\label{fig:klee_diagram}
\end{figure}

\begin{figure}[htbp]
  \centering
  \includegraphics[height=0.65\linewidth]{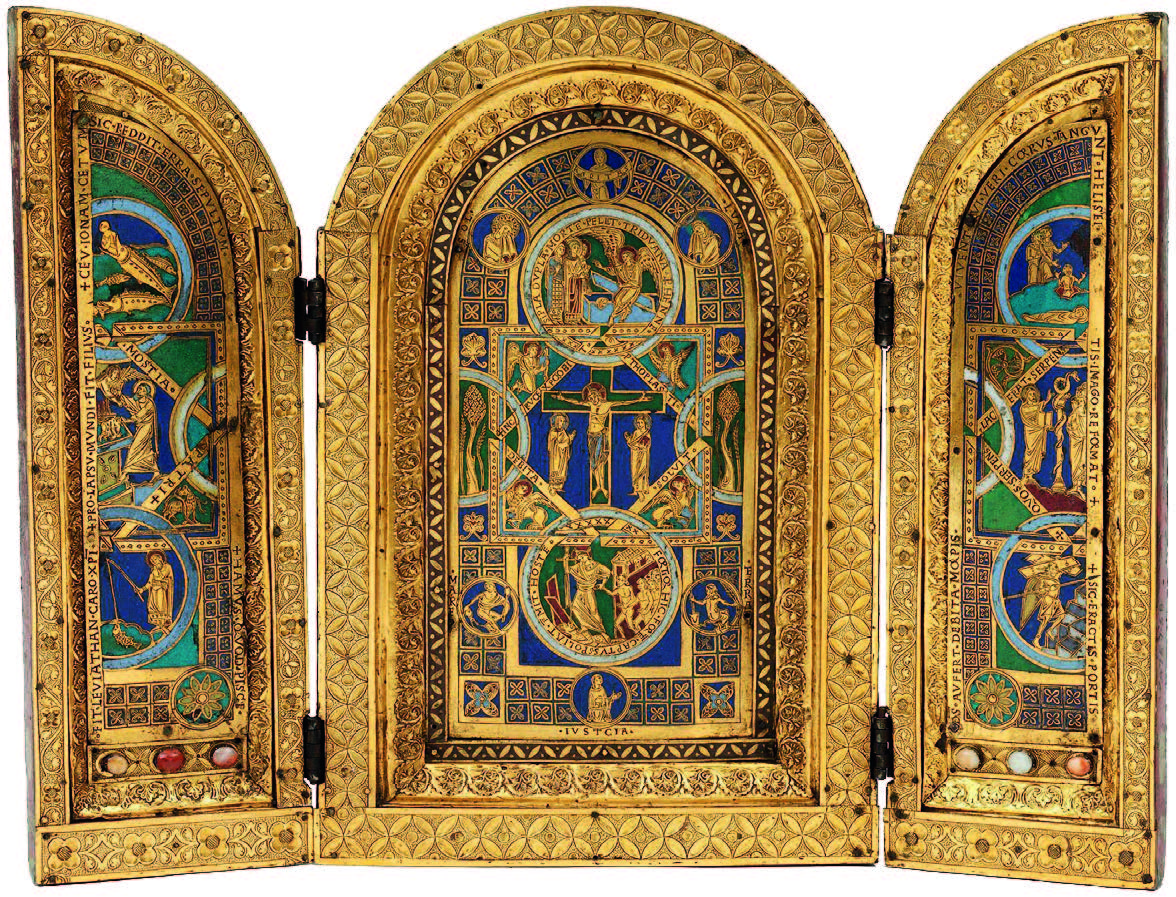}
  \caption{An example of diagram in religious art taken from \cite{hamburger2019devotion}: Alton Towers Triptych, Cologne (?), ca. 1150. London, Victoria \& Albert Museum, inv. no. 4757-1858. Photo: © Victoria \& Albert Museum.}\label{fig:triptych_diagram}
\end{figure}

\section{Background and Related Works}

In his work, Peirce often emphasized the role of diagrams in reasoning~\cite{dewaal2013perplexed}, providing various examples of what he termed \textit{diagrammatic reasoning} (CP 4.571, 5148, 6.213). On the one hand, diagrams exhibit an iconic character: they are a specific subtype of icon capable of representing the internal structure of the objects they depict through the arrangement of their interconnected parts. For example, a map that shows historical trade routes by positioning ports and drawing connecting lines already functions as a diagram, as it represents spatial and relational structures that mirror real-world networks of movement and exchange. On the other hand, diagrams also possess a dynamic character, as they enable manipulation and iterative transformations according to the general laws governing the relationships among their parts, and as such, pose epistemological questions about the generation and production of knowledge~\cite{stjernfelt2007diagrammatology,kiryushchenko2023diagrams}. For instance, a simple triangle drawn on paper can serve as a diagram of all triangles by altering its angles or side lengths while preserving its topology. Most notably, Peirce developed a system of visual logic known as\textit{ Existential Graphs}, structured into four levels: \textit{Alpha} for propositional logic, \textit{Beta} for modal and higher-order logic, and \textit{Gamma} and \textit{Delta} for meta-assertions and non-declarative statements. \textit{Existential Graphs }use visual connectors such as lines, curves, and nodes to represent logical relations. To start, the \textit{Sheet of Assertion} is a blank space on which diagrams are drawn. Any proposition written directly on it is considered to be asserted as true. A continuous line connecting the elements indicates a logical identity. Enclosures, such as closed curves, represent negation; therefore, a region inside a curve is logically negated. Juxtaposed elements without connectors are logically conjunctive (i.e., both are assumed to be true). A bifurcation indicates a logical disjunction. These characteristics make Peirce’s diagrams interesting for computational modeling; however, their formal and visual complexity also requires methods capable of isolating and structuring heterogeneous visual content within manuscripts.

The structural heterogeneity of historical manuscripts poses significant challenges for automated information extraction and semantic indexing. Recent studies have developed machine learning pipelines for layout segmentation, targeting both textual and non-textual elements, such as illuminations and decorative initials~\cite{aouinti2022illumination, buttner2022cordeep}. Object detection models, such as YOLO~\cite{yaseen2024yolov8indepthexplorationinternal}, have also been employed to identify image regions in complex manuscript layouts, offering effective layout segmentation~\cite{ravichandra2022deep}. Among these approaches, Fleischhacker et al.~\cite{fleischhacker2025ocr} proposed a pipeline that combines layout detection, synthetic data augmentation, OCR fine-tuning, and reintegration of outputs as IIIF-compliant annotations. While such methods enable the scalable processing of visually rich documents, they do not incorporate mechanisms for semantic enrichment or integration into LOD frameworks. To address this issue, ontological extensions of the Web Annotation Data Model (WADM)~\cite{w3c_web_annotation} have been proposed. The Multi-Level Annotation Ontology (MLAO)~\cite{pedretti2024mlao} introduces conceptual anchoring and provenance for annotations, while the General Ekphrastic Ontology (GEkO)~\cite{bocchi2025gekphrastic} models ekphrastic relations between textual descriptions and visual elements.
These limitations also highlight the need for approaches that combine visual segmentation with semantic interpretation. In this context, VLMs represent promising tools for bridging the gap between image analysis and knowledge extraction processes.

In recent years, the intersection of computer vision and art history has seen a growing integration of visual and textual modalities, driven by the availability of large art collections of digitized images and the development of multimodal machine learning techniques. This has led to substantial interest in tasks such as visual link retrieval, multimodal classification, iconographic captioning~\cite{cetinic2021iconclass}, and visual question answering (VQA)~\cite{garcia2018readings}, particularly in domains such as cultural heritage, where image content is often accompanied by curatorial or scholarly metadata. Although multimodal models can describe visual elements, they often struggle to understand the logical or spatial relationships among them. This limitation has been highlighted in recent studies that show how VLMs tend to rely on background knowledge rather than analyzing the internal structure of diagrams~\cite{hou2024vlmvisual}. To overcome this, some approaches combine image segmentation and structured prompting. For example, the chain-of-regions method decomposes diagrams into meaningful areas before interpreting them, improving the model’s reasoning about spatial relations~\cite{zhang2024improvevisionlanguagemodel}. Other studies have applied VLMs to structured visual domains, such as UML diagrams or flowcharts, using modular reasoning pipelines to achieve more accurate results~\cite{liang2024flowlearn}. However, to the best of the authors' knowledge, no existing study has specifically addressed the use of VLMs for identifying or interpreting diagrammatic content in historical sources. In this context, applying similar strategies to Peirce’s existential graphs means using layout segmentation to isolate diagram regions and then prompting VLMs with questions informed by Peirce’s semiotics. This structured workflow helps models provide more accurate interpretations of diagrammatic reasoning.

\section{Methods} 

\subsection{Corpus Description and Preprocessing}

The Charles S. Peirce Papers (MS Am 1632)~\cite{Robin1971-ROBTPP-2}, housed at Harvard’s Houghton Library, represent one of the most extensive archival collections of Peirce’s works. Comprising over 1,700 manuscript items, the collection spans disciplines ranging from mathematics to logic and metaphysics. A subset of 233 items was digitized and made available through IIIF Manifests via the Harvard Hollis system~\cite{harvard2023peirce}, yielding a total of 15,695 high-resolution facsimile images.

To prepare the corpus for computational processing, we retrieved IIIF metadata for each digitized item, including canvas structure, image URIs, and classification labels derived from Robin’s catalogue~\cite{Robin1971-ROBTPP-2}. All canvases were downloaded at full resolution and organized into thematic folders. Blank pages, identified using IIIF metadata, were automatically excluded, resulting in a set of 13,234 manuscript pages (Table~\ref{tab:corpus-summary}).

\begin{figure}[H]
\centering
\includegraphics[width=0.9\linewidth]{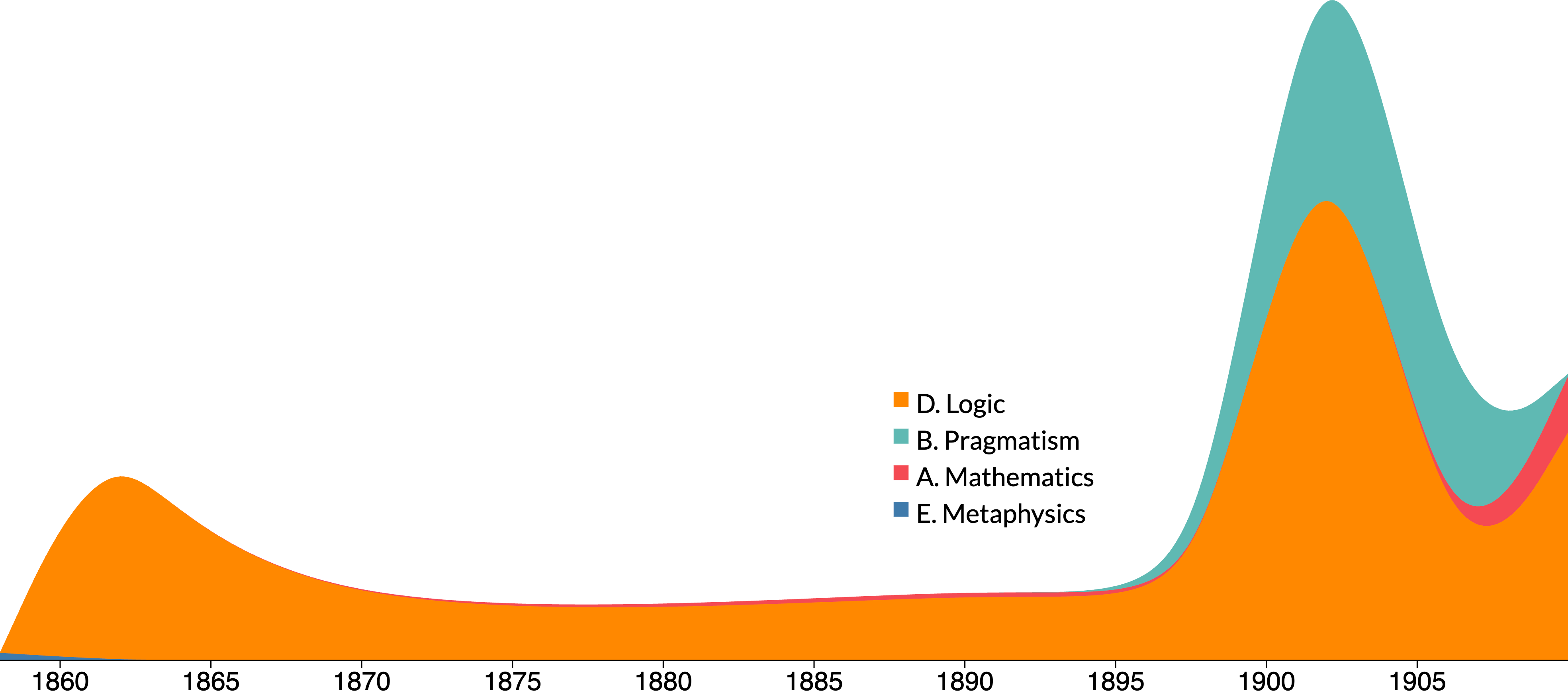}
\caption{Distribution of digitized manuscript pages across Peirce’s lifetime, grouped by five-year intervals and categorized according to Robin’s classification. The visualization, based on IIIF canvas data, highlights how Peirce’s intellectual focus evolved over time, with Logic manuscripts dominating the corpus and peaks corresponding to his most productive years in formal reasoning.}\label{fig:bumpchart-pagecount}
\end{figure}

To contextualize the corpus thematically, we constructed a bump chart showing the distribution of digitized pages in Robin’s topical categories grouped by five-year intervals within Peirce’s lifetime (Figure~\ref{fig:bumpchart-pagecount}). Category D (Logic) dominates the corpus, followed by Pragmatism (B) and Metaphysics (E), reflecting Peirce’s focus on formal reasoning and motivating our attention to visual content in these areas.

\begin{table}[H]
\centering
\begin{tabular}{l r}
\toprule
\textbf{Description} & \textbf{Count} \\
\midrule
Total manuscript items           & 1,759 \\
Digitized items                  & 233 \\
Total digitized pages            & 15,695 \\
Blank pages removed              & 2,461 \\
Pages retained for processing    & 13,234 \\
\bottomrule
\end{tabular}
\caption{Key statistics of the Peirce manuscript corpus.}\label{tab:corpus-summary}
\end{table}

\subsection{Models for Page and Layout Analysis}

To distinguish textual from visually mixed pages, we implemented a classification pipeline based on three feature extraction strategies: Histogram of Oriented Gradients (HOG), intermediate features from ResNet18, and semantic embeddings from the CLIP visual encoder. Each page was labeled as either \textit{text} (pages containing mostly textual elements), \textit{diagram\_mixed} (pages containing at least one relevant visual feature), or \textit{cover} using a manually annotated dataset of 1,264 pages.

To extract diagrammatic content from the classified \textit{diagram\_mixed} pages, we created a manually annotated dataset of 443 manuscript images. Each page was labeled using a two-class schema: \texttt{diagram} and \texttt{text\_block}. This typology was designed to identify the main visual and textual regions while preserving the layout-level structure and supporting generalization across diverse manuscript formats. At this stage, we intentionally avoided adding more granular annotations, such as dates, titles, sketches, logical notation, or algebraic formulas, to prevent class imbalances in the training data.

The dataset was split into training and validation subsets using an 80/20 ratio. To address the class imbalance and increase robustness, we applied data augmentation to all pages containing at least one diagram annotation. Two synthetic variants were generated for each page, resulting in a final training set of 1,133 images. Finally, we fine-tuned the YOLOv8m model on this dataset, which was selected for its balance between detection performance and computational efficiency.

\subsection{Annotation Workflow}
All detected segments, whether diagrams or text blocks, were transformed into structured annotations compliant with the IIIF. Each segment is expressed as an instance of WADM, linking a specific region of the manuscript image (the \texttt{target}) to a content resource or metadata element (the \texttt{body}), serialized in JSON-LD format. Bounding boxes are mapped to IIIF Canvas coordinates using the \texttt{xywh} fragment selector, ensuring the precise anchoring of visual elements within the page layout.

To enhance semantic expressiveness, we employ MLAO~\cite{pedretti2024mlao}, an extension of WADM. The MLAO introduces the \texttt{mlao:Anchor} class to separate the annotated physical region from its conceptual referent. In our use case, anchors are linked to the IIIF URI of the full manuscript page via \texttt{mlao:isAnchoredTo}, enabling a shared conceptual reference for both textual and diagrammatic segments of the manuscript. Instead of predefined abstraction layers (e.g., Work, Expression, Manifestation, and Item according to LRMoo~\cite{riva2022lrmoo}), we define custom conceptual categories based on Peirce’s semiotic theory (see \S\ref{sec:vlm_prompting}). Interpretative captions generated by the VLM are modeled as \texttt{oa:TextualBody} instances linked to a \texttt{hico:InterpretationAct}~\cite{daquino2015hico} that specifies the interpretative level, model used, and generation process via PROV-O. This structure supports the hermeneutic traceability and versioning of automated interpretations.

Annotations are generated from the detection outputs and can be embedded in IIIF Manifests or published as standalone pages. Annotations can also be serialized in RDF for semantic querying. This semantic layer supports integration into LOD workflows and prepares the content for VLM interpretation and use.

\subsection{VLM Prompting and Interpretation}\label{sec:vlm_prompting}

Peirce’s semiotic theory offers a framework to understand the structure and function of signs, which we use as a basis to design VLMs prompts. We define three analytical categories for prompt engineering that operationalize aspects of Peirce's semiotic theory for computational analysis. These categories are designed for VLM prompting rather than direct applications of his icon-index-symbol trichotomy. The morphological level addresses the basic visual elements that constitute a diagram, such as lines, shapes, and symbols. This corresponds broadly to the iconic mode of representation and relates to Peirce's category of Firstness. The indexical level concerns the relationships between these elements, identifying connections, dependencies, or structural links. This reflects aspects of Peirce's notion of Secondness. The symbolic level explores the logical operations encoded in the diagram. At this stage, we provide the VLM with minimal instructions on how to interpret visual conventions (e.g., enclosure, juxtaposition, lines of identity), prompting the model to reconstruct the inferential logic underlying the structure. This aligns with Peirce's category of Thirdness. Table~\ref{tab:vlm-question-templates} shows the specific question templates for each category.

Finally, we defined three classes extending the MLAO data model (\texttt{pip:MorphologicalLevel}, \texttt{pip:IndexicalLevel}, and \texttt{pip:SymbolicLevel}) to represent the semiotic categories described earlier. The VLM-generated responses, along with metadata about the model and prompt, were re-injected as annotations into the IIIF-compliant JSON-LD structure. Each annotation targets a specific region of the IIIF canvas and references the full manuscript page via its persistent URI. 

\begin{table}[ht]
\centering
\begin{tabular}{lp{11cm}}
\toprule
\textbf{Semiotic Level} & \textbf{Question Template} \\
\midrule
\textbf{Morphological} & How many and what kind of elements (e.g., words, lines, arcs, nodes, shapes, etc.) are present in the image? \\
\textbf{Indexical} & Is there a relationship between the elements present in the image? Which elements are connected to each other? \\
\textbf{Symbolic} & In Peirce’s diagrammatic logic, a closed curve called a \textit{cut} represents logical negation. 
Elements inside the same region are interpreted conjunctively (i.e., asserted together).
Elements placed directly on the background (the Sheet of Assertion) are considered true.
A cut around propositions denies them. Nested cuts represent nested negation.
Lines may indicate identity or existential quantification.

Based on these principles, interpret the diagram and translate its meaning into a logical statement. If this is not possible, provide a clear explanation in natural language. \\
\bottomrule
\end{tabular}
\caption{Template of VLM Questions Based on Semiotic Categories}\label{tab:vlm-question-templates}
\end{table}

\subsection{Evaluation Methodology}

To assess the interpretative capabilities of VLMs with respect to Peirce’s diagrammatic logic, we conducted a qualitative evaluation across five diagrams of increasing complexity, manually selected from the Peirce manuscript corpus and belonging to the Alpha level. Standard reference-based metrics, such as CLIPScore~\cite{hessel2021clipscore} are limited in this context, as they primarily measure lexical similarity and do not account for the semantic or structural accuracy of a caption~\cite{darmenio2025semiotic, cetinic2018fineart}. This makes them unsuitable for evaluating the descriptions of abstract and diagrammatic content. For each diagram, three structured prompts were submitted to the models, corresponding to Peirce’s semiotic categories: morphological (element enumeration), indexical (relational structure), and symbolic (logical translation), as shown in Table~\ref{tab:vlm-question-templates}. We tested five VLMs: \textbf{BLIP3-o}~\cite{Chen2025BLIP3}, \textbf{GPT-4o}, \textbf{LLaVA 1.6 vicuna-13b}~\cite{liu2023visualinstructiontuning}, \textbf{MiniGPT-4 vicuna-13b}~\cite{zhu2023minigpt4enhancingvisionlanguageunderstanding} and \textbf{Phi-4 Multimodal}~\cite{haider2024phi3safetyposttrainingaligning}, all of which accept both visual and textual inputs. This evaluation aimed to compare their capacity to recover structured meaning from diagrammatic forms, with particular attention to inferential depth and semiotic coherence.

Each response was rated on a 3-point scale: 2 for correct and complete answers, 1 for partially correct answers, and 0 for incorrect or irrelevant responses. The total possible score was 30 (3 questions × 5 diagrams × 2 points).

\section{Results and Discussion}

\subsection{Performance of Preparatory Models}

\begin{figure}[ht]
    \centering
    \includegraphics[width=0.9\textwidth]{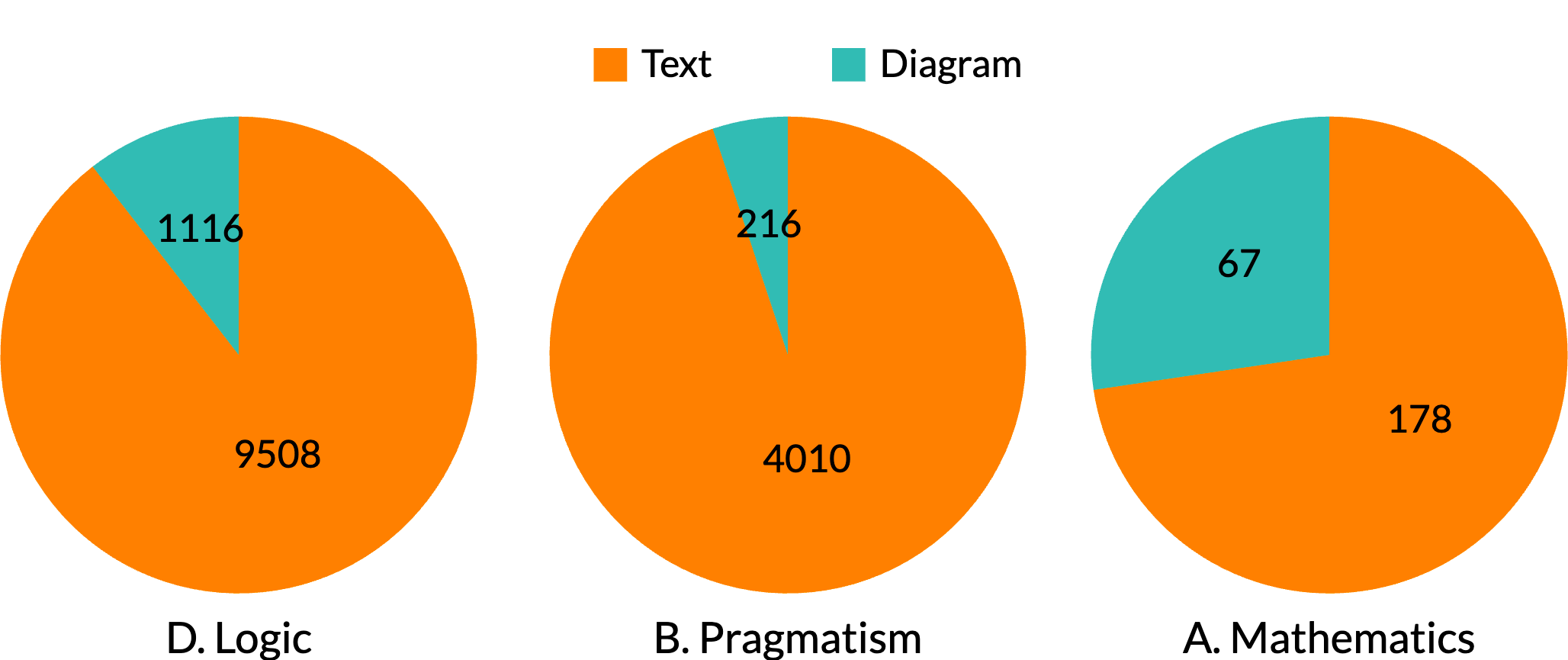}
    \caption{Pie charts show text and diagrams distribution within three categories.}\label{fig:diagram-vs-text}
\end{figure}

The best performance for image classification was achieved using a logistic regression classifier trained on CLIP embeddings. Using 10-fold stratified cross-validation, the model achieved a macro-averaged F1-score of \textbf{0.9531}, with good class-wise accuracy across the board. The full results of the model comparison are reported in Appendix~\ref{appendix:classification}. To assess the distribution of visual content across the corpus, we applied the trained classifier to all digitized pages and aggregated the predictions according to thematic categories. As shown in Figure~\ref{fig:diagram-vs-text}, visual content is especially concentrated in Category D (Logic), followed by Pragmatism (B) and Metaphysics (E), suggesting that Peirce used visual reasoning more frequently in these areas.

\begin{table}[ht]
\centering
\begin{tabular}{lrrrr}
\toprule
Class & Precision & Recall & F1 Score & Support \\ \midrule
Cover (0) & 1.0000 & 1.0000 & 1.0000 & 28 \\
Text (1) & 0.9459 & 0.8974 & 0.9211 & 117 \\
Diagram (2) & 0.9040 & 0.9496 & 0.9262 & 119 \\
\bottomrule
\end{tabular}
\caption{Performance of the best model (Logistic Regression + CLIP) by class.}\label{tab:logistic-regression}
\end{table}

\begin{table}[ht]
\centering
\begin{tabular}{lccc}
\toprule
\textbf{Metric} & \textbf{Diagram} & \textbf{Text Block} & \textbf{All Classes} \\
\midrule
mAP@0.5 & 0.992 & 0.970 & 0.981 \\
Precision (at best F1) & 0.990 & 0.960 & 0.975 \\
Recall (at best F1) & 1.000 & 0.940 & 0.990 \\
Optimal F1 score & 0.996 & 0.948 & 0.960 \\
Confidence @ Optimal F1 & 0.547 & 0.547 & 0.547 \\
Confidence @ Max Precision & 0.975 & 0.975 & 0.975 \\
Confusion (TP) & 684 & 473 & – \\
Confusion (FP) & 38 & 58 & – \\
Confusion (FN) & 3 & 23 & – \\
\bottomrule
\end{tabular}
\caption{Detection performance on the Peirce manuscript validation set (n=111 images).}\label{tab:metrics}
\end{table}

\begin{figure}[hbtp]
  \centering
  \includegraphics[width=0.73\linewidth]{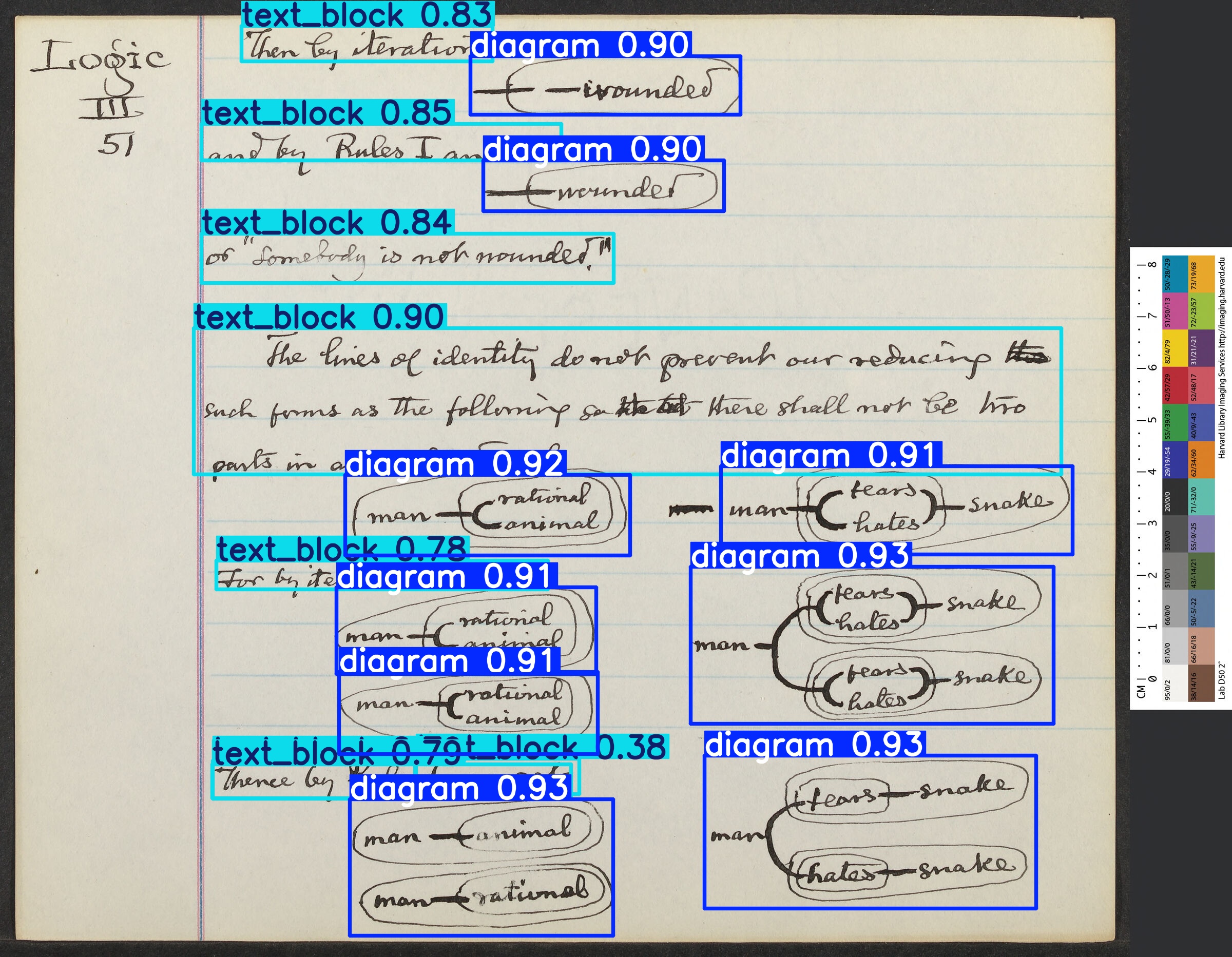}
  \caption{Output of the fine-tuned YOLOv8m model on autograph manuscript dated 1902. MS Am 1632 (430), Box 29, Folder 28, Series I. Manuscripts, D. Logic. Houghton Library, Harvard University, USA. Photo: © Houghton Library. Persistent URL: \url{http://nrs.harvard.edu/urn-3:FHCL.HOUGH:12491033}. The model correctly identifies and segments 'diagram' (blue) and 'text\_block' (light blue) regions, providing the structured data used for subsequent annotation and VLM analysis.}\label{fig:annotated-peirce}
\end{figure}

On the validation set (111 pages), the fine-tuned YOLOv8m model achieved a mean Average Precision at IoU 0.5 (mAP@0.5) of \textbf{0.981}, with a class-specific score of \textbf{0.992} for \texttt{diagram} and \textbf{0.970} for \texttt{text\_block}. The precision, recall, and F1 scores are reported in Table~\ref{tab:metrics}. A sample prediction with annotations is shown in Figure~\ref{fig:annotated-peirce}.\footnote{The models and the scripts used for the preprocessing and evaluation are available at: \url{https://anonymous.4open.science/r/PIP-Manuscripts-Processor-0147/}.}

\subsection{Preliminary VLM Evaluation}

As shown in Table~\ref{tab:qualitative_scores}, GPT-4o achieved the highest score (25/30), demonstrating relatively strong performance across all semiotic dimensions. BLIP3-o followed with 12/30, showing partial competence in recognizing visual elements but struggling with relational and symbolic interpretation. Phi-4 obtained a total of 6 points, with modest success in symbolic recognition but weak results in the other dimensions. Both LLaVA 1.6 and MiniGPT-4 scored 0, failing to produce meaningful responses to any of the evaluative questions.

\begin{figure}[H]
  \centering
  \includegraphics[width=0.4\linewidth]{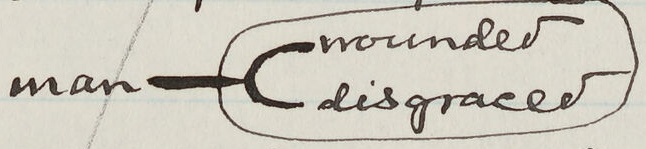}
  \caption{Example diagram from Peirce's \emph{Logic} manuscripts, used for evaluating the interpretative capacity of VLMs.}
  \label{fig:peirce-diagram-example}
\end{figure}

While GPT-4o and BLIP3-o can handle layout-level tasks without fine-tuning, their performance drops significantly when confronted with more complex diagrams involving nested cuts or non-trivial spatial configurations. Some errors are attributable to OCR-like misrecognition, such as reading “wounded” as “mound” or “man” as “noun”. Across nearly all models, the symbolic level obtained the lowest average scores, with frequent failures in understanding negation correctly (e.g., misinterpreting a cut as emphasis, ignoring it entirely) and generating logically valid formalizations (e.g., confusing \( \lnot (A \land B) \) with \( \lnot A \land \lnot B \)), or even hallucinating logical rules (in smaller models). Interestingly, considering the diagram in Figure~\ref{fig:peirce-diagram-example}, both GPT-4o and BLIP3-o produced formal logical statements in response to the symbolic question. GPT-4o correctly generated: “There exists a man who is not both wounded and disgraced,” and formalized it as:
\begin{align}
\exists x \left( \text{Man}(x) \land \lnot \left( \text{Wounded}(x) \land \text{Disgraced}(x) \right) \right). \label{eq:gpt4o}
\end{align}

BLIP3-o instead produced: “It is not the case that there exists a man who is wounded and disgraced,” rendered as:

\begin{align}
\lnot \exists x \left( \text{Man}(x) \land \text{Wounded}(x) \land \text{Disgraced}(x) \right). \label{eq:blip3o}
\end{align}

This suggests that while BLIP3-o demonstrates surface-level competence in formalization, GPT-4o is better able to align visual features with underlying logical relations.

To finalize the workflow suggested at the end of \S\ref{sec:vlm_prompting}, we also produced the RDF serialization of the annotations generated by the model.\footnote{The dataset and RDF serializations are available at \url{https://doi.org/10.5281/zenodo.16113285}.}

\begin{table}[ht]
\centering
\begin{tabular}{lccc|c}
\toprule
\textbf{Model} & \textbf{Morphological} & \textbf{Indexical} & \textbf{Symbolic} & \textbf{Total} \\
\midrule
GPT-4o         & 7                     & 9                  & 9                 & 25             \\
BLIP3-o        & 3                     & 5                  & 4                 & 12             \\
Phi-4          & 1                     & 1                  & 4                 & 6              \\
LLaVA 1.6      & 0                     & 0                  & 0                 & 0              \\
MiniGPT-4      & 0                     & 0                  & 0                 & 0              \\
\bottomrule
\end{tabular}
\caption{Qualitative evaluation scores across five diagrams. Maximum: 30 points per model.}\label{tab:qualitative_scores}
\end{table}

The high concentration of diagrams in Peirce's manuscripts offers a quantitative overview of the importance of visual representations as working instruments during reasoning itself, aligning with the philosopher pragmatic maxim that concepts acquire meaning through their operational use rather than from correspondence to abstract or \textit{a priori} definitions. In other words, Peirce recurred to diagrammatic reasoning on paper as a privileged way to develop his arguments and to enable the manipulation of relational structures. Through this perspective, this exploratory analysis reveals Peirce's philosophy as an embodied practice where "moving pictures of thought" and linear writing operate in a continuous exchange, underlining at the same time the importance of intermediality in such heterogeneous manuscripts, supporting the arguments raised by \textcite{keeler2020a_hidden}. Our preliminary VLM evaluation serves as a pilot study. Expanding the dataset would enable systematic investigation of how Peirce's model of semiosis relates to how VLMs process diagrammatic content, and whether studying these models can in turn clarify what diagrammatic reasoning requires. The workflow presented here makes such evaluation feasible across the full corpus.

\section{Conclusion}

This preliminary work presents a modular pipeline for analyzing heterogeneous manuscript collections, combining layout classification, object detection, and semantic annotation within IIIF and LOD frameworks. The workflow extends WADM through MLAO and introduces a qualitative VLM evaluation method structured around analytical categories derived from Peirce's semiotic theory. Applied to Peirce's manuscripts, the analysis reveals the quantitative distribution shows that visual reasoning is concentrated in specific philosophical domains, with Logic manuscripts containing 10.5\% diagrams versus 5.1\% in Pragmatism. This shows that diagrammatic practice was functionally integrated into Peirce's work on formal systems. Second, VLM evaluation reveals that GPT-4o can approximate logical interpretation when appropriately prompted (25/30 points), while smaller models underperform. The differential performance across morphological, indexical, and symbolic questions validates these as functionally distinct analytical operations. 

The methodological pattern extends beyond Peirce studies. Any manuscript collection combining text and visual elements can adapt this workflow by substituting domain-appropriate theoretical frameworks, retraining segmentation models, and adjusting prompts to specific research questions. Moreover, the workflow can be integrated into semantic digital editions based on digital facsimiles. Textual content can be transcribed using HTR, while annotations contextualize visual elements at multiple scales. The annotations generated through this process can automatically enrich knowledge graphs. Finally, by treating annotations as digital traces of interpretation, the system aligns with Peirce's pragmatist view of semiosis as an ongoing process.


\printbibliography

\appendix

\section{Model Comparison for Page Classification}
\label{appendix:classification}

\begin{table}[ht]
\centering
\label{tab:classification-summary}
\begin{tabular}{llrrrr}
\toprule
Feature & Model & Avg Precision & Avg Recall & Avg F1 Score & Accuracy \\ \midrule
CLIP & Linear SVM & 0.9335 & 0.9320 & 0.9321 & 0.9091 \\
CLIP & Logistic Regression & \textbf{0.9500} & \textbf{0.9490} & \textbf{0.9491} & \textbf{0.9318} \\
CLIP & Random Forest & 0.9296 & 0.9293 & 0.9294 & 0.9053 \\
CLIP & SVM RBF & 0.9500 & 0.9460 & 0.9461 & 0.9280 \\
CLIP & k-NN (k=5) & 0.9123 & 0.9093 & 0.9093 & 0.8788 \\
CNN & Linear SVM & 0.9351 & 0.9350 & 0.9350 & 0.9129 \\
CNN & Logistic Regression & 0.9356 & 0.9349 & 0.9350 & 0.9129 \\
CNN & Random Forest & 0.9183 & 0.9180 & 0.9181 & 0.8902 \\
CNN & SVM RBF & 0.9183 & 0.9180 & 0.9181 & 0.8902 \\
CNN & k-NN (k=5) & 0.8932 & 0.8835 & 0.8882 & 0.8561 \\
HOG & Linear SVM & 0.8341 & 0.8243 & 0.8290 & 0.7765 \\
HOG & Logistic Regression & 0.8361 & 0.8361 & 0.8361 & 0.7803 \\
HOG & Random Forest & 0.8688 & 0.8375 & 0.8504 & 0.8182 \\
HOG & SVM RBF & 0.8548 & 0.8351 & 0.8442 & 0.8030 \\
HOG & k-NN (k=5) & 0.6659 & 0.6974 & 0.5889 & 0.6212 \\
\bottomrule
\end{tabular}
\caption{Performance summary of all models across features. Best values are in bold.}
\end{table}

\begin{table}[ht]
\centering
\label{tab:confusion-matrix-appendix}
\begin{tabular}{lccc}
\toprule
\textbf{True / Predicted} & Cover & Text & Diagram\_mixed \\
\midrule
Cover           & 28  & 0   & 0   \\
Text            & 0   & 105 & 12  \\
Diagram\_mixed  & 0   & 6   & 113 \\
\bottomrule
\end{tabular}
\caption{Confusion matrix for Logistic Regression + CLIP embeddings (aggregated over 10 folds).}
\end{table}

\end{document}